\def\final{1}
\documentclass[fleqn,10pt]{wlscirep}
\usepackage[utf8]{inputenc}
\usepackage[T1]{fontenc}
\usepackage{lineno}
\usepackage{orcidlink}

\usepackage{graphicx}
\usepackage{nameref}
\usepackage{float}
\restylefloat{figure}

\usepackage{cleveref}
\crefformat{equation}{(#2#1#3)}
\crefformat{figure}{Fig.\,#2#1#3}
\crefformat{section}{Sec.\,#2#1#3}
\crefformat{appendix}{App.\,#2#1#3}
\crefformat{table}{Tab.\,#2#1#3}

\newcommand{\secref}[1]{``\nameref{#1}''}

\makeatletter
\newcommand\setcurrentname[1]{\def\@currentlabelname{#1}}
\makeatother

\usepackage[final]{changes}

\definecolor{JV}{RGB}{0,150,255}
\definecolor{MK}{RGB}{150,0,150}
\definecolor{HM}{RGB}{100,0,200}
\definecolor{TT}{RGB}{0,150,0}
\definecolor{JS}{RGB}{250,0,200}
\definecolor{ST}{RGB}{200,0,0}
\definecolor{MDe}{rgb}{0.0, 0.48, 0.65}
\definechangesauthor[color=JV]{JV}
\definechangesauthor[color=MK]{MK}
\definechangesauthor[color=HM]{HM}
\definechangesauthor[color=TT]{TT}
\definechangesauthor[color=JS]{JS}
\definechangesauthor[color=ST]{ST}
\definechangesauthor[color=MDe]{MDe}

\makeatletter
\@namedef{Changes@AuthorColor}{red}
\colorlet{Changes@Color}{red}
\makeatother

\ifx\final\undefined
\newcommand{\note}[2][]{\added[#1]{\footnotesize\textit[#2]}}
\else
\newcommand{\note}[2][]{}
\fi

\usepackage{tabularx}

\usepackage{listings}

\lstdefinestyle{mystyle}{
    basicstyle=\ttfamily\footnotesize,
    breakatwhitespace=false,
    breaklines=true,
    captionpos=b,
    keepspaces=true,
    showspaces=false,
    showstringspaces=false,
    showtabs=false,
    tabsize=2
}

\usepackage{xurl}

\title{Metadata practices for simulation workflows}

\author[ \orcidlink{0009-0007-8791-7100} 1,2,*]{José Villamar}
\author[ \orcidlink{0000-0001-9303-6712} 3]{Matthias Kelbling}
\author[ \orcidlink{0000-0002-7514-2199} 1,4]{Heather L. More}
\author[ \orcidlink{0000-0003-1255-7300} 1]{Michael Denker}
\author[ \orcidlink{0000-0001-5794-5425} 1]{Tom Tetzlaff}
\author[ \orcidlink{0000-0002-6304-062X} 1,5]{Johanna Senk}
\author[ \orcidlink{0000-0003-3939-1523} 3]{Stephan Thober}
\affil[1]{Institute for Advanced Simulation (IAS-6), Jülich Research Centre, Jülich, Germany}
\affil[2]{RWTH Aachen University, Aachen, Germany}
\affil[3]{Department of Computational Hydrosystems, Helmholtz-Centre for Environmental Research, Leipzig, Germany}
\affil[4]{Institute for Advanced Simulation (IAS-9), Jülich Research Centre, Jülich, Germany}
\affil[5]{Sussex AI, School of Engineering and Informatics, University of Sussex, Brighton, United Kingdom}
\affil[*]{corresponding author: José Villamar (j.villamar@fz-juelich.de)}

\begin{abstract}
  Computer simulations are an essential pillar of knowledge generation in science.
  Exploring, understanding, reproducing, and sharing the results of simulations relies on tracking and organizing the metadata describing the numerical experiments.
The models used to understand real-world systems, and the computational machinery required to simulate them, are typically complex, and produce large amounts of heterogeneous metadata.
Here, we present general practices for acquiring and handling metadata that are agnostic to software and hardware, and highly flexible for the user.
These consist of two steps: 1) recording and storing raw metadata, and 2) selecting and structuring metadata.
As a proof of concept, we develop the \emph{Archivist}, a Python tool to help with the second step, and use it to apply our practices to distinct high-performance computing use cases from neuroscience and hydrology.
Our practices and the \emph{Archivist} can readily be applied to existing workflows without the need for substantial restructuring.
They support sustainable numerical workflows, fostering replicability, reproducibility,
data exploration, and data sharing
in simulation-based research.
\end{abstract}

\begin{document}

\maketitle
\thispagestyle{empty}

\phantomsection
\section*{Introduction}
\setcurrentname{Introduction}
\label{sec:intro}
Recent advances in high-performance computing (HPC) technology enable simulations of increasingly large and complex models.
While these simulations offer huge potential for science and society, it becomes more and more challenging to replicate and reproduce the results of such numerical experiments, to efficiently explore the simulation data, and to share them with other scientists.

User stories such as the following are therefore common among researchers:
\begin{enumerate}
\item \emph{\textbf{Replicating results:} 
    Scientist X cannot replicate the results of scientist Y due to inconsistencies between the information provided in the scientific article and in the associated code published by Y.
    Even personal communication with Y does not resolve these inconsistencies\cite{Pauli18}.}
\item \emph{\textbf{Data sharing:} 
    Each member of a group of scientists regularly runs simulations of the same mathematical model to investigate different scientific questions.
    The datasets generated by each of these scientists are similar and potentially useful to other members of the group, although what one scientist considers to be metadata can be analyzed as primary data by another scientist with a different objective.
    It is therefore desirable to share these data to minimize time and energy costs and ultimately to increase scientific productivity.
    However, the scientists have no efficient way of communicating the information necessary to understand the structure of each dataset\cite{Leipzig2021}.}
\item \emph{\textbf{Data exploration:} 
    A group of scientists is developing a simulation software.
    After each development cycle, the group runs a set of benchmarking experiments with different configurations and models to continuously monitor the simulation software performance (``continuous benchmarking''\cite{Antz19}).
    After years of development, the group has accumulated large amounts of benchmarking data for each software version.
    The scientists have no way to easily search and view subsets of the accumulated data, for example, to compare different versions using similar configurations or model types.}

\end{enumerate}
\par
One would assume that the above problems, known to different extents in various research fields, should not exist in simulation science, due to our perception of full control over digital implementations\cite{Penders2019}.
However, this assumption is typically wrong for two reasons.
First, users are often not aware of every aspect of their hardware and software systems, such as low-level hardware settings\cite{Gutzen18_90}, implementation details of a software package, or the implications of using a specific operating system\cite{Glatard15_12}.
These aspects are of particular importance for HPC-enabled simulations, where highly specialized hardware and software solutions and the distribution of code among multiple processors may affect the exact simulation outcome or performance measures. 
Given the limited lifetime of HPC systems, long-term repeatability is
practically impaired.
Second, users often customize aspects of their system without properly documenting their changes, for example modifying the behavior of off-the-shelf software.
Therefore, despite the digital nature of simulation research, it remains difficult to acquire and organize metadata describing the details of hardware systems, software stacks, model descriptions and simulation workflows that are necessary to replicate, reproduce, understand, and share the results of numerical experiments\cite{Nordlie-2009_e1000456,Ivie2018,Pauli18,McDougal16_2021}.
These metadata arise at all stages of the simulation workflow, including defining and implementing a model; preparing software; generating and executing a job; post-processing, analyzing, and visualizing data; and organizing and storing data.

\par

\par

Throughout this paper, we use the terms `reproducibility' and `replicability' as defined by the Association for Computing Machinery (ACM) (\url{https://www.acm.org/publications/policies/artifact-review-and-badging-current}, last access: 12 February 2025).
According to these guidelines, results from a computational experiment are \emph{reproduced} if an independent team obtains the same results using the same experimental setup, whereas results are \emph{replicated} if an independent team obtains the same results using a different experimental setup.
Here, `experimental setup' refers to the artifacts used to generate results, which ``can be software systems, scripts used to run experiments, input datasets, raw data collected in the experiment, or scripts used to analyze results''.
For simulation experiments, we consider such artifacts to be simulation engines, model and workflow implementations, and their respective configurations.
However, it is important to consider not only the software, but also the hardware system used to obtain results.
For instance, performance metrics such as time to solution or energy consumption cannot be interpreted without knowledge of the hardware on which the experiment is run.
These considerations may blur the conceptual line between reproduction and replication.
It is, for example, not always clear whether a small change in the configuration or the hardware leads to a different experimental setup.
Ultimately, it is the research question underlying a simulation experiment that determines which aspects of the experimental setup are important to consider when reproducing or replicating results.

\par

Successfully capturing comprehensive metadata and provenance information describing the experiment and its setup is a prerequisite for reproducibility and replicability\cite{Nordlie-2009_e1000456,Davison12_48,Ivie2018,Manninen2018,Plesser18_76,McDougal16_2021}, allows assessment of simulation outcomes\cite{Guilyardi2013,Pauli18}, and helps scientists explore, share, and reuse data\cite{Albers22_837549}.

Recently, several powerful tools have been developed to organize, execute, and track complex workflows, including their resulting data and metadata, such as Sumatra\cite{Davison12_48}, AiiDA\cite{Huber20_4}, Snakemake\cite{Mlder2021}, and DataLad\cite{Halchenko2021}.
Although these tools are useful for new projects, using them in existing projects with different, possibly custom, workflow management systems requires deciding whether to rebuild the workflow from scratch using the desired tools or look for alternative solutions..
As a consequence, scientists often write custom code for metadata handling to compromise between adding functionality and keeping an established workflow.
The resulting inconsistency in metadata management makes it difficult to transfer knowledge between researchers.
\par
Other community efforts have focused on defining vocabularies or schemas such as RO-Crate\cite{SoilandReyes2022}, CodeMeta (\url{https://codemeta.github.io}, last visited: 23 January 2025), or Bioschemas\cite{bioschemas}.
Such standards enable interoperability of data and metadata produced by computational workflows, and provide practical ways for developers and users to describe and document their experiments.
Nonetheless, detailed information on individual workflow executions can only be collected during runtime and often requires additional transformations to comply with vocabularies or schemas.
Examples of such information are: the software environment at runtime, computing capabilities of the hardware system in use, or performance and resource consumption of computationally heavy workflow steps.
Furthermore, identifying where such information can be found and the methods to transform it depend on the workflow implementation and system used.
This dependency makes found solutions harder to transfer to other workflows or systems, and consequently researchers avoid collecting and handling such information unless required by their scientific objectives. 
\par
Here, we propose a set of practices for flexible metadata management which can readily be integrated into existing simulation workflows without substantial refactoring.
The practices apply to commonly used workflows across diverse research fields.
They advise researchers about which metadata to collect at each stage of a simulation workflow, how to collect it, and how to process the metadata so that they enrich data.
We further describe our Python tool \emph{Archivist} which helps process metadata files using user-defined functions and combine the outputs into a unified file.
Finally, we apply our proposed practices and the \emph{Archivist} to a minimal illustrating example, and to two real-world use cases from neuroscience and hydrology.

\phantomsection
\section*{Results}
\setcurrentname{Results}
\label{sec:Results}

\phantomsection
\subsection*{Metadata management practices}
\setcurrentname{Metadata management practices}
\label{sec:practices}

\begin{figure}[!htb]
  \centering \includegraphics[width=\textwidth]{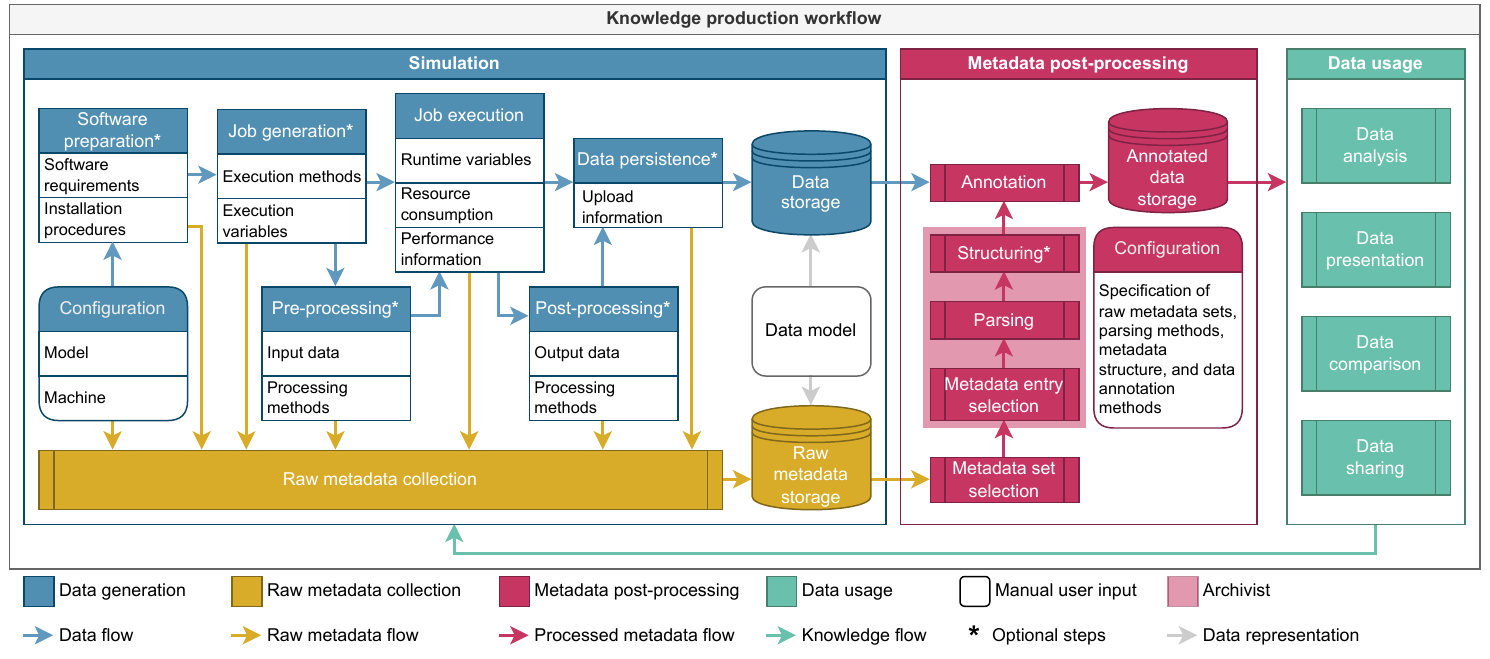}
  \caption{
    \textbf{Metadata management in a generic knowledge production workflow.}
    We conceptually divide the workflow into three consecutive sub-workflows: simulation, metadata post-processing, and data usage.
    \textbf{Simulation} collects raw metadata (yellow) at each step (blue), then stores it according to a data model describing the structures of data and raw metadata.
    \textbf{Metadata post-processing} (red) selects raw metadata, then parses and structures it; annotation links this processed metadata to the data.
    During post-processing, user-defined configuration specifies which raw metadata to select and how to process it, ensuring the final annotated data is suitable for its application.
    Some of these steps (light red) can be performed by the post-processing framework presented in~\secref{sec:archivist}.
    \textbf{Data usage} (green) analyzes data to draw conclusions through presentation and comparison; data may also be shared with others.
    Ultimately, the results of data usage inform subsequent simulations (green arrow).
    Each sub-workflow consists of autonomous steps (rectangles), user-configured steps (rounded rectangles), data storage (cylinders), and data transfer (arrows).
    This diagram aims to abstract simulation-based research; specific workflows may omit some elements (asterisks) or have additional steps.
}
  \label{fig:workflow}
\end{figure}

To better understand the metadata collection and handling processes, we consider a generic knowledge production workflow divided into three sub-workflows (\cref{fig:workflow}): an abstract simulation experiment containing processes generating data and metadata, metadata post-processing where heterogeneous metadata is processed into meaningful formats and structures, and usage of enriched data.
Although the implementation of these processes can and will vary across different use cases, this generic knowledge production workflow encompasses archetypes of the components occurring in many implemented workflows.

An abstract simulation experiment comprises a sequence of steps (\cref{fig:workflow}, \emph{Simulation} section).
We define these steps by summarizing the processes performed in two domain-specific workflows from neuroscience and hydrology\cite{Albers22_837549,Thober2019}.
We consider the necessary steps to be configuration of the software environment, simulation engine, and model; execution of the simulation; and storage of the generated data and metadata; depending on the complexity of the preparation required before or after the execution of the simulation, there might be additional steps.
Note that we do not claim that this abstract simulation experiment exhaustively represents all workflows or use cases.

After a simulation experiment, users can analyze the resulting data and metadata to gain knowledge of the simulated model or simulation technology.
The methods through which data is analyzed are tightly related to the objectives and intentions of the user.
Furthermore, in some disciplines, these methods may require a complex workflow and perhaps even the use of HPC resources.
Since we cannot exhaustively define the different existing strategies for processing data, we provide an abstract description of how the generated data can be exploited (\cref{fig:workflow}, \emph{Data usage} section).
For instance, data can be analyzed in order to measure metrics related to model predictions or accuracy, or simulation performance.
Graphical visualizations are generated for easier human interpretation.
Additionally, differences between multiple versions of simulators, models or parameters can be obtained by comparing with previous results.
Then, newly obtained insights can be shared with peers to disseminate knowledge.

Because
processing the data may inadvertently remove or add specific features that alter the original information, storing the original data together with a data model allows independent users to exploit individual datasets with different methods, or to process multiple datasets at once, which increases re-usability of the simulation results. Over time, the data and metadata generated by different simulation experiments can be compiled and stored in a continuously growing collection.
Establishing a data model is necessary to maintain the organization and integrity of this collection.
A data model is an abstract representation of data objects and their relationships that organizes relevant attributes of data and standardizes how they relate to each other\cite{Alagić1986}.
This serves as a blueprint for structuring data elements and their relationships, providing a clear understanding of how data is organized, stored, and utilized within a collection.
Without the data model, users may not be able to establish an overview of the actual contents and may fail to efficiently explore the collection.
For clarity we separate data and metadata storage in \cref{fig:workflow}, however this is an implementation choice left to the discretion of the user and for the remainder of this paper we refer to both storages together as one collection for simplicity. 

To implement and perform the diversity of data usage tasks, metadata is required to properly describe the simulation experiment.
When exploring multiple sets of generated data, metadata categorizing each dataset allows researchers to efficiently differentiate them.
Moreover, when sharing the results, metadata enhances the knowledge transfer between data users by detailing what, how, and where the experiment was performed.
Metadata can exist at any step of the simulation (\cref{fig:workflow}, blue \emph{Data generation} boxes), and can often be accessed only when performing the experiment.
The collection of metadata and their interpretation often varies and evolves with the systems and software used (\cref{fig:workflow}, yellow \emph{Raw metadata collection} boxes).
We describe below some general practices for successfully collecting and handling the variety of metadata that arises during simulation (\cref{fig:workflow}, \emph{Metadata post-processing} section).

\paragraph{Collect as much information as possible.} 
Metadata are data that describe data.
Here, we focus on information that can be directly collected in the environment where the experiment takes place (sometimes termed "hard" metadata \cite{Grewe11_16}).
However, the definition of what precise pieces of information should be represented as metadata is difficult to generalize.
Each experiment generates an arbitrary amount of information -- separating this information into metadata and data depends on the nature and goal of the workflow itself.
For instance, if the goal of a simulation workflow is to collect and analyze the data generated by the underlying model, then the performance of the simulation (time to solution, memory consumption, etc.) would be part of the metadata.
On the contrary, if the goal of the workflow is to benchmark the simulator then the data generated by the model is of secondary importance, as the recorded performance data would be the desired primary result.
Limiting the amount of information collected during a simulation experiment restricts how its output can be used in the future.
Therefore, to maximize the potential for reuse of shared experiment results, it is important to collect as much information as possible\cite{Davison12_48}.
Yet, the full extent of this collection needs to be determined on a case by case basis.
Sometimes, collecting the full breadth of information available can be redundant.
For example, when simulation models grow complex, the number of available parameters grows in volume too.
Well-documented parameter sets often use default values suitable for their corresponding model; during experiments, only a subset of these may actually be modified.
In this case, collecting only the modified parameters may be sufficient.
Conversely, there may be instances where keeping detailed track of the differences between experiment is necessary.
Sometimes, differences in software versions  lead to discrepancies in experiment results\cite{Pauli18}.
In the case of rapidly evolving software such as the model, simulator, or their library dependencies, it is beneficial to keep track of the different versions used for each experiment.
As a compromise between efficiency and completeness, we propose collecting information as sustainably as possible to an extent where all currently known user needs are satisfied.

\paragraph{Identify metadata sources.} Given the rapidly evolving software and hardware platforms available for simulation alongside user-specific set-ups, it is impossible to exhaustively list all collectible metadata.
Instead, we categorize the different sources of metadata according to each step of the simulation experiment (\cref{fig:workflow}, blue \textit{Data generation} boxes):
\begin{itemize}

    \item \emph{Configuration}: Most steps during the workflow need some kind of configuration or parametrization, which the user often saves and edits in files beforehand.
    These files are a type of metadata describing each step of the simulation. This metadata includes, in particular, the configuration and parameters used by the model and simulator.

    \item \emph{Software preparation}: In this step all the required software (libraries, tools, and programs) are fetched, compiled if needed, and loaded into the environment.
    Information to collect here includes the source of the software, the parameters related to compilation, and the environment variables defined when loading the software.

    \item \emph{Job generation} : In HPC scenarios, job schedulers are generally used to efficiently allocate and share resources of compute clusters.
    Collectable metadata here include job parameters like machine name and partition used, amount and type of resources used, resource configuration such as process binding or thread pinning, and job time limit.

  \item \emph{Pre-processing}: Some simulation experiments require input data to calibrate or parametrize the model implementation.
      Using this input data may involve additional calculations or transformations within the workflow.
    If pre-processing is necessary, documenting the input data and procedures performed on it must be recorded to accurately determine the starting point of the simulation. 

    \item \emph{Job execution}: The execution is the main part of the simulation, and involves instantiating the simulation model and running the simulation.
    It is at this step that recording the status of the execution environment  is the most informative and least redundant.
    In contrast to \emph{Software preparation}, the metadata collected in this step should focus on the environment variables, libraries, and processes present during execution.
    Additionally, performance information, resource consumption (memory, storage, network, etc.), and system load can only be recorded at this step.

    \item \emph{Post-processing} : Depending on the goal and implementation of the workflow, the data that is generated must be post-processed for further use, which may involve steps such as checking, cleaning, compressing, or transforming the data.
    If post-processing is necessary, documenting the procedures performed on data is essential to have an overview of how the final result was obtained. Thus information about expected output and the different methods used during this step must be recorded.

    \item \emph{Data persistence}: In this step, information on the storing process can be recorded. For example: who generated the data and metadata to be stored? When was it generated and when is it being stored? Where was it generated and where is it being stored e.g. a file server, a local or remote database?
    In the case of a file server, the address of the server and the absolute path of the stored files should be recorded.
    When using a database, new entries are indexed with a unique identifier; storing this identifier allows linking with future related entries and should be recorded correspondingly.
\end{itemize}

Depending on the system and software used, the metadata present in the above mentioned categories can often be found as parameter or configuration files.
In particular, most system information can be obtained through commands provided by the operating system or package management solutions.

Users might be able to recognize the metadata required for a specific data analysis and collect it alongside other metadata they are aware of.
However, situations could arise where reusing the generated data for unforeseen purposes is impossible due to missing context.
Even after collecting all known metadata, it may be that there are still unknown pieces of metadata that need to be recorded.
With rapidly evolving systems and software, this is not an uncommon scenario, and if such a situation arises then users must identify what information is lacking,  and add it to the list of metadata to be recorded for future experiments.

\paragraph{Metadata collection does not need to be complicated.} There are two possible ways of collecting metadata: retrieval by a dedicated parallel process “pulling” the relevant pieces of information from each simulation step, or autonomous “pushing” by each component to a metadata storage.
The pull-based approach relies on an overarching metadata-monitoring module that is adapted to the specific simulation.
The push-based approach, in contrast, does not require such infrastructure at run time.
At each step, the governing process simply dumps its metadata in some arbitrary, more or less raw format, independent of other simulation steps.
The push-based approach is therefore much more flexible and can readily be applied to existing simulations.
With each component presenting its own metadata it is likely that the final collection of metadata contains different structures and formats.
Although it is possible to standardize the presentation of metadata across all components, this process would require transformations that are dependent on the objective for which the data is being generated, and can vary from user to user, or the interoperability standard that the researcher must adhere to.
In some situations, transformations of metadata may even depend on other metadata, making it difficult to perform an ad-hoc transformation process.
We therefore propose extracting and structuring the raw metadata in a separate step after the simulation.
This presents a clear advantage by decoupling metadata collection processes from user goals and processing software.
Collecting as much metadata as possible is essential as it is impossible to get the corresponding metadata after a simulation is performed.
Thus, gathering the metadata regardless of format and storing it during simulation for later processing is an efficient way to keep track of the information describing the experiment.
We call the collection of these unprocessed metadata files the ``raw" metadata (\cref{fig:workflow}, yellow \emph{Raw metadata collection} boxes).

\paragraph{Post-process recorded metadata according to your goals.} Transforming the heterogeneous raw metadata into a unified comprehensive format is essential in order to be able to exploit it alongside the data.
Given that each workflow is run with a specific objective in mind, the collection of raw metadata often contains much more than needed.
Although this is desirable to maximize possible reusability of simulation data, for a given objective the raw metadata should be filtered.
For this, a dedicated read operation that extracts the relevant entries from a collection of metadata is performed according to the given objective.
The extracted metadata is parsed and might be immediately usable depending on the output format of the parsing operations, however a specific structure might be needed to increase interoperability between the processed metadata and other data exploitation software.
Therefore, structuring the metadata according to a standard is often a beneficial step to easily and unambiguously leverage the metadata information.
With this, the previously heterogeneous metadata collection is now unified, possibly structured according to a standard, and ready to be exploited.

\paragraph{Combine your data with metadata.} Before being able to use processed metadata with data  need to be combined . Here, we refer to combining as a referencing procedure in which any data entity that makes up a complete data record is cross-referenced with any metadata (item or collection) that assists in using and interpreting the data entity.
Although it is possible to exploit metadata and data independently for specific purposes, combining them quickly becomes necessary when the volume and dimension of a dataset start to grow.
When amassing data from multiple simulations, each with varying degrees of different configuration and/or models, it becomes apparent that simulation results need to be described at a higher level .
For this, ``tagging" or annotating the data with the processed metadata creates an enriched dataset.
When appropriately stored, users can exploit indexing engines to navigate through enriched datasets and generate complex data projections.
Again, the level of granularity with which data entities are identified and matched to metadata will need to be decided based on the objective of the workflow.

\medskip

Following the above practices will help in collecting and handling large heterogeneous volumes of metadata; additionally, combining datasets with processed metadata increases their explorability and exploitability.
The actual methods through which these practices can be followed or implemented depend on the technology used to manage the simulation experiment and data processing pipeline.
In particular some existing technologies offer support for these practices and are addressed in the \secref{sec:discussion}.

\phantomsection
\subsection*{Metadata post-processing framework -- the \emph{Archivist}}
\setcurrentname{Metadata post-processing framework -- the \emph{Archivist}}
\label{sec:archivist}

Users who follow the metadata collection practices outlined in the previous section might end up with large volumes of heterogeneous metadata.
For example, users could potentially collect: code; documentation; system information; human readable and non-human readable data; binary, encoded, and compressed files; etc...
Immediately exploiting the information found in this collection might be practically unfeasible due to the variety of file structure and formats.
In particular, database systems and analysis tools often require specific input structures or formats without much flexibility.
To be able to use their collected metadata, users need to individually transform the information in each file and then combine it into a specific output, or look for available software to aid in this process.

Although there exist tools capable of doing these kinds of operations, e.g. AiiDA\cite{Huber20_4} parsers (\url{https://.readthedocs.io/projects/-core/en/latest/reference/apidoc/.parsers.html}, last access: 23 February 2024), and DataLad\cite{Halchenko2021} metadata extractors (\url{https://docs.datalad.org/projects/metalad/en/latest/extractors.html}, last access: 23 February 2024), they may constrain how metadata is processed,  and/or may be an inseparable component of a larger framework without the ability to be used independently.

Here, we propose the \emph{Archivist} as a framework for performing parsing and structuring operations.
Our goal is to facilitate the implementation of our metadata handling practices by providing the \emph{Archivist} as an intermediary tool to extract information from a heterogeneous metadata file collection and unify it into a single file.
Some users will want to structure the unified metadata file according to a vocabulary or schema such as RO-Crate\cite{SoilandReyes2022}, CodeMeta, or Bioschemas\cite{bioschemas}.
Whereas our proposed metadata practices aim to collect information describing the conditions under which an individual result was obtained, vocabularies and schemas provide a higher level of description for entities, context, and goals surrounding the experiment itself.
Nonetheless, some of the collected metadata can help to generate documents according to these defined standards. To do this, users can define data templates to structure the extracted information from the collection of metadata files into a comprehensive output compliant with their desired standard.
Still, users will likely need to provide additional input to fully exploit the expressivity of the vocabularies.
An example output compliant with the Bioschemas vocabulary for computational workflows is shown in \secref{sec:methods}.

The \emph{Archivist} doesn't currently support data annotation, i.e., linking metadata records to relevant parts of the data in the post-processing pipeline (\cref{fig:workflow}, red \emph{Annotation} box), because the heavy customizations required to fit the underlying structure of the data are difficult to formalize.
For example, while data structured according to HDF5-based\cite{Koranne2010} formats like the NIX\cite{Adrian14_} standard can make use of built-in mechanisms for linking collections of metadata entries to data, such mechanisms cannot necessarily be generalized to other file formats and data models.

\begin{figure}[!htb]
  \centering
  \includegraphics[width=\textwidth]{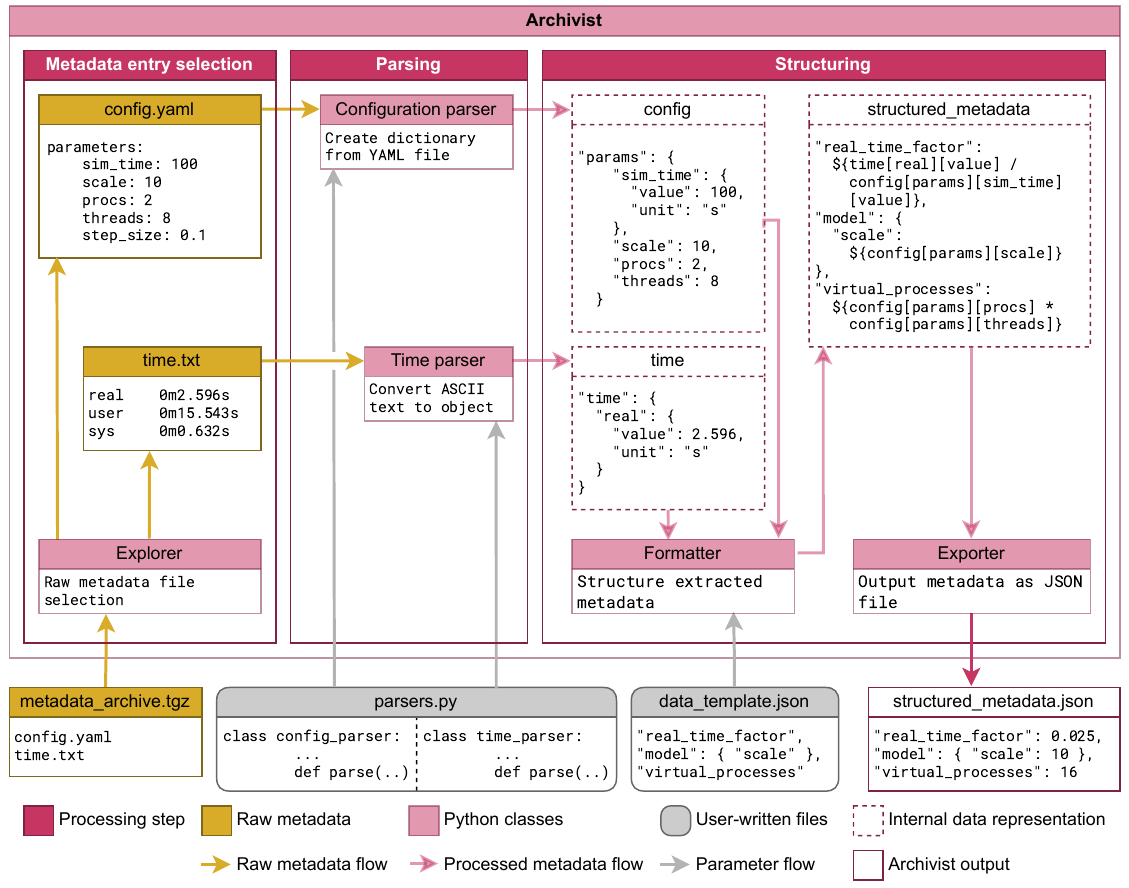}
  \caption{
    {\bf Implementation of an \emph{Archivist} metadata processing pipeline with two example parsers.}
    From the user perspective, the \emph{Archivist} class is provided with all inputs associated with the pipeline (bottom yellow and grey boxes).
    Internally, the \emph{Explorer} class extracts the individual files to process from a collection of raw metadata files, and dispatches them to corresponding \emph{Parser} classes, here the \textit{Configuration parser} and the \textit{Time parser} classes.
    Then, each \emph{Parser} class employs a user-defined function to extract specific information from its respective files.
    After this, the \emph{Formatter} class collects the parsing results. If a data template is provided, the composite result can also be restructured. The final processed metadata is output in a format of choice by the \emph{Exporter} class.
      A Jupyter tutorial explaining how to implement the metadata processing pipeline shown here is available at \url{https://github.com/INM-6/metadata-archivist/blob/main/examples/schema_example4/schema_tutorial.ipynb} (last access: 26 January 2025). 
  }
  \label{fig:archivist}

\end{figure}

To illustrate the \emph{Archivist} framework, we show a representative example implementation of a workflow to parse and structure metadata in \cref{fig:archivist}.
The framework, coded in Python (\url{http://www.python.org}, last access: 23 February 2024), is primarily composed of four processing classes and one interface class.
The processing classes perform different aspects of the metadata parsing and structuring (exploration of metadata files, parsing of metadata sources, formatting and structuring of collected metadata, and exporting to a selected format), while the central interface class orchestrates the other classes .

Below, we introduce the functionalities of each of these classes.
Source code, example implementations, and additional details can be found at \url{https://doi.org/10.5281/zenodo.13442425} (last access: 30 August 2024).

\paragraph{Archivist class:} The \textit{Archivist} class is a convenience interface class which instantiates and orchestrates the processing classes.
As input, the class accepts a collection of raw metadata, the parsing operations to apply, and optionally a data template.
This interface is necessary because each processing class uses specifically structured inputs and outputs, which can be cumbersome and error-prone for users to generate themselves.
We designed the interface to be configurable, giving it enough flexibility to customize the behavior of the processing classes while providing simple inputs and outputs.

\paragraph{Explorer class:} The \textit{Explorer} class processes inputs given to the \emph{Archivist}.
These can be raw metadata archives or directories containing raw metadata files.
To enhance flexibility, no assumption is made about the structure and contents of the archives or directories. As such, the user must define rules identifying which files to parse -- we refer to this as a file target rule.
To do this, the user can provide precise file names or regular expressions describing these names.
We refer to these as file description rules.
Using these rules, the \emph{Explorer} searches the input (archive or directory) for corresponding files and provides a list of files to parse.

\paragraph{Parser abstract class:} The \textit{Parser} is an abstract classed designed to be extended by users to extract metadata from files.
When instantiating a \textit{Parser}, the user must associate one of the file target rules provided to the \textit{Explorer} .
Depending on the rule, the \textit{Archivist} instance dispatches each file to the respective \textit{Parser}.

Additionally each user must provide their own parsing methods.
Two examples of such instances are shown in \cref{fig:archivist}: the \texttt{config\_parser} which employs the PyYAML (\url{https://github.com/yaml/pyyaml}, last visited: 26 January 2025) package to read YAML files, and the \texttt{time\_parser} which uses a custom ASCII file reader. The ASCII reader consists of a simple line-by-line parsing loop over the input file, using a regular expression to select only lines starting with a word followed by a space then a time marker composed of minutes and seconds.
If the word matches the string ``real", then the time marker is stored and its time value converted to seconds.
The final output is a dictionary containing the value in seconds of the real time and a unit description. 
Additional examples of parsing classes are provided with our source code and we hope to build a user base around shared methods to foster reusability and replicability.

\paragraph{Formatter class:} The \textit{Formatter} combines the output of the Parser instances into a unified metadata file.
Although parsing the desired metadata files and listing the results in a single file might be sufficient for some workflows, in other cases it may be necessary to transform the collection of parsing results into a cohesive and comprehensive structure.
For this, the user must provide a data template to match a parsing result to the desired structure.
These data templates are defined with an extended implementation of the JSON Schema\cite{Pezoa2016} which the \textit{Formatter} can interpret to link data structures with extracted information.
As a simplified example in \cref{fig:archivist}, given the parsed information from the configuration file and time file, the user can combine the simulation time with the real time in a single field as the \texttt{real\_time\_factor} .

\paragraph{Exporter class:} The \textit{Exporter} saves the internal data representation of the structured metadata to a file with specific output format.
Like the \textit{Parser}, though not an abstract class itself, this class was designed to be extensible to enable serialization to arbitrary formats.
In particular, users define the output format that enhances compatibility with their annotation method of choice.
In \cref{fig:archivist}, the \emph{Archivist} employs the JSON format to export the structured metadata.

\medskip

Using this class hierarchy, users only need to define the parsing functionalities they are interested in along with the file target rules, and provide a metadata collection to process.
The \textit{Archivist} class takes care of coordinating the other classes and generating the output.
For more complex operations, users can provide a data template for the \textit{Formatter} class, and extend the \textit{Exporter} class to change the output format.
With this flexibility, the \emph{Archivist} framework provides a re-usable parsing and structuring pipeline that can operate on existing metadata or be attached to a workflow for automated post-processing.
A Jupyter tutorial explaining how to implement the metadata processing pipeline shown in \cref{fig:archivist} is available at \url{https://github.com/INM-6/metadata-archivist/blob/main/examples/schema_example4/schema_tutorial.ipynb} (last access: 26 January 2025).

\phantomsection
\subsection*{Examples}
\setcurrentname{Examples}
\label{sec:examples}
To illustrate how the proposed metadata practices (\secref{sec:practices}) and the \emph{Archivist} framework (\secref{sec:archivist}) can be implemented into specific simulation workflows, we provide here one minimal and two real-world examples.
The first example (\secref{sec:minimal_example}) constitutes a toy workflow applicable to simulations where the user conducts a parameter scanning experiment and leverages recorded metadata to identify suitable configurations.
The second example (\secref{sec:neuroscience_example}) describes a workflow for the benchmarking and the verification of a specific neuroscience simulation architecture.
The third example (\secref{sec:hydro_examples}) showcases a routine procedure for the calibration of a hydrological model.

\phantomsection
\subsubsection*{Minimal example}
\begin{figure}[!htb]
  \centering
  \includegraphics[width=\textwidth]{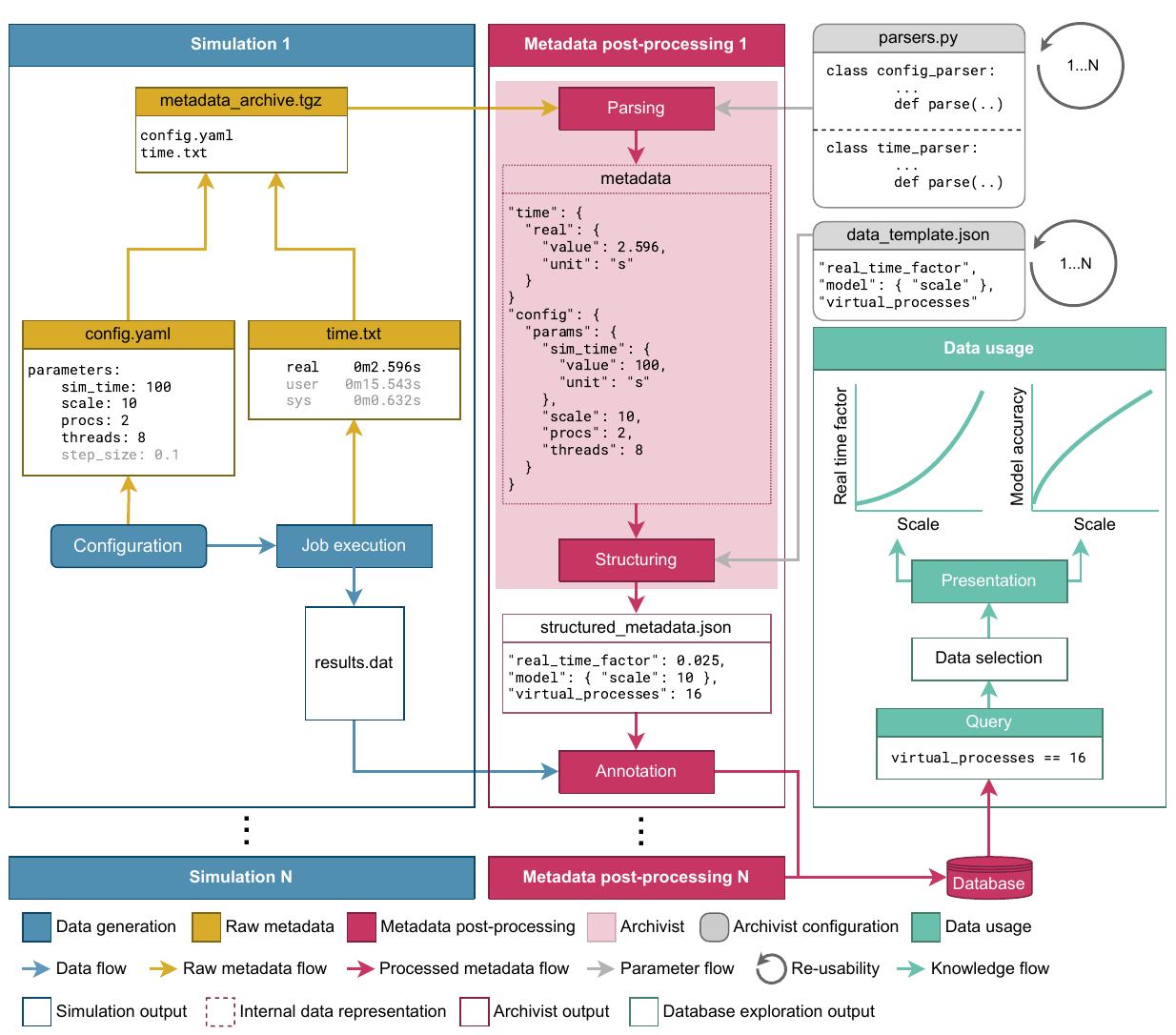}
  \caption{
    {\bf Minimal example.}
    Illustration of the \emph{Archivist}'s functionality in a simple example use case.
    In a parameter scanning experiment, several instances of a model with different configurations (parameters) are simulated (``Simulation $1$'',\ldots, ``Simulation $N$''; blue boxes on the left).
    During each simulation, configuration and performance information are recorded and stored in a (raw) metadata archive (yellow).
    After each simulation, the stored metadata is post-processed (``Metadata post-processing $1$'',\ldots, ``Metadata post-processing $N$''; red):
    first, the relevant information is extracted by user-defined \emph{Parser} classes (gray box \texttt{parsers.py}).
    Non-relevant information is discarded (see light gray text in the  raw metadata files).
    The extracted metadata are then structured according to a provided data template (gray box \texttt{data\_template.json}).
    Finally, the simulation results are annotated with the structured metadata and stored in a database (red cylinder).
    After all simulation and  metadata post-processing instances are finished and their corresponding results are stored in the database, the annotated data can be queried and presented (green).
}
  \label{fig:mini_example}
\end{figure}

\setcurrentname{Minimal example}
\label{sec:minimal_example}
The first example addresses a typical question in computational science (\cref{fig:mini_example}): how can we choose the parameters of a given model such that the accuracy of its predictions is high while the time-to-solution is small?
In many applications, a typical parameter which improves the model prediction but slows down the simulation is the model size, here referred to as the \texttt{scale}.
Examples are the number of elements in a finite element simulation, or the number of neurons in an artificial neuronal network.
The minimal example described here illustrates how the enrichment of simulation results with processed metadata helps find an answer to the above question.
For simplicity, we assume that the example runs in a local simulation environment where the required software stack is already installed, no job manager is used, and the data is stored locally. 
In accordance with \cref{fig:workflow}, we subdivide the entire workflow into the three components: \emph{Simulation}, \emph{Metadata post-processing}, and \emph{Data usage}.
Below, we describe each of these components in detail.
\par
The  \emph{Simulation} section consists of the configuration of the model and the simulation architecture, the simulation execution, and the collection of the generated data and metadata (\cref{fig:mini_example}, blue box).
Here, we do not further specify the actual structure and dynamics of the underlying hypothetical model. 
We merely assume that it contains a parameter \texttt{scale} representing the model size. 
During the simulation, the model dynamics are propagated forward in time with some time resolution (\texttt{step\_size}), up to a total model time (\texttt{sim\_time}).
We further assume that the simulation architecture is equipped with a parallelization infrastructure, which is parameterized by the total number of processes (\texttt{procs}) and the number of computing threads used by each process (\texttt{threads}).
Model and architecture parameters are stored in a configuration file (\texttt{config.yaml}).
During each simulation, some primary simulation data (which we do not further specify here) is generated and stored in a data file (\texttt{results.dat}).
In addition, the simulation scripts monitor the duration of each simulation run with the help of the Linux \textit{time} command (\url{https://www.man7.org/linux/man-pages/man1/time.1.html}, last access: 25 August 2024).
It returns three different types of durations: the \texttt{real} time represents the actual wall-clock time, i.e., the total time taken by the simulation process from invocation to termination, while the \texttt{user} and the \texttt{sys} times measure the cumulative CPU (thread) time spent in the user and in the kernel mode, respectively.
All three times (\texttt{real}, \texttt{user}, \texttt{sys}) are stored in a text file (\texttt{time.txt}).
The simulation is re-run for a range of \texttt{scale}'s, \texttt{procs}, and \texttt{threads}.
At the end of each simulation run, the configuration and the time files are collected in the form of a raw metadata archive (\texttt{metadata\_archive.tgz}).
\par
Within the \emph{Metadata post-processing} section, relevant configuration information and simulation times are parsed from the raw metadata archive and combined into a structured metadata file (\cref{fig:mini_example}, red box).
For the question at hand, not all metadata collected during the simulation needs to be extracted.
Let's, for example, assume that the time resolution (\texttt{step\_size}) was kept constant for all simulations.
Moreover, we are only interested in the \texttt{real} but not in the \texttt{user} and in the \texttt{sys} times.
During the parsing of the raw metadata, only a subset of metadata is therefore extracted (black items in \texttt{config.yaml} and \texttt{time.txt} in \cref{fig:mini_example}).
In addition, physical units are added to the time quantities.
The metadata extraction is performed by the \emph{Archivist} with the help of user-defined parsing classes (\texttt{parsers.py}).
The \emph{Archivist} further structures the extracted metadata according to a user-defined template (\texttt{data\_template.json}).
In this example, the template introduces the total number of virtual processes (\texttt{virtual\_processes}) calculated by multiplying the number of processes by the number of threads per process, as well as the real time factor (\texttt{real\_time\_factor}) calculated by taking the ratio between the measured wall-clock (\texttt{real}) time and the model time (\texttt{sim\_time}).
Finally, the primary simulation data are annotated by the structured metadata.
In this example, annotation is performed by uploading the primary simulation data and the corresponding structured metadata as a single entry to a local database.
An implementation of the metadata post-processing performed for this example can be found at \url{https://doi.org/10.5281/zenodo.13442425} (last access: 30 August 2024).
\par
In the \emph{Data usage} section, the local database containing the accumulated annotated simulation results is queried to ultimately find model sizes where the model accuracy is sufficiently high while the time-to-solution is low (\cref{fig:mini_example}, green box).
To this end, the parameter search is restricted to a subset of simulations where a given amount of computational ressources have been used.
In this example, only those entries are extracted from the database where the total number of virtual processes equals $16$.
Based on the selected primary simulation data, the user assesses the model accuracy, for example, by comparing the model predictions with some observed or experimental data (for a real-world example, see \secref{sec:hydro_examples}).
At the same time, the user extracts the real time factor from the corresponding structured metadata.
A subsequent analysis of the dependence of the model accuracy and the real time factor on the model size (\texttt{scale}), and an account of additional constraints such as the maximum acceptable real time factor or the minimal acceptable model accuracy, permits an identification of appropriate model sizes.

\phantomsection
\subsubsection*{Neuroscience use case}
\setcurrentname{Neuroscience use case}
\label{sec:neuroscience_example}
Understanding how the brain ``computes'', what principles it employs to solve complex tasks with minimal energy consumption, how it evolves and changes during the lifetime of an organism, what the origins and effects of neurodegenerative diseases are, and what possibilities of treatment exist has huge social, economical, and ecological impact.
The human brain consists of about $10^{11}$ nerve cells (neurons)\cite{Herculano-Houzel09}, which form a complex network.  
Each neuron receives inputs from thousands of other neurons, both from its local neighborhood and from distant brain areas.
The connectivity structure is highly heterogeneous and depends on the involved neuron types and brain areas.
Furthermore, the connections between neurons (synapses) are not static but change depending on sensory inputs and other factors.
The mathematical description of the brain's dynamics at cellular resolution therefore involves large sets of coupled differential equations.
Even for a single cubic millimeter of brain tissue, this number is on the order of at least $10^4$.
Neuroscience is thus dealing with complicated mesoscopic dynamical systems, which are neither small nor in the thermodynamic limit.
They can not be fully understood by means of analytical mathematical methods from dynamical systems theory or statistical physics.
State-of-the-art neuroscience hence relies on simulation.
\par  
Simulating brain-scale neuronal networks at cellular resolution, i.e., instantiating the corresponding models and simulating them in a reasonable time, is challenging.
In particular, investigating slow biological processes, such as learning or brain development, requires accelerated simulations where the wall-clock times $T_\text{wall}$ are substantially smaller than the duration $T_\text{model}$ of the simulated time interval. 
Due to the large number of neurons and connections, brain-scale neuronal-network simulation requires substantial amounts of memory to store all involved state variables\cite{Kunkel2012_5_35}.
At each instance of time, large numbers of differential equations have to be solved simultaneously.
Brain simulation thus typically employs parallel, distributed computing.
One of the key challenges in distributed brain simulation, however, is the communication between neurons, which are typically located on different compute nodes.
Developing efficient algorithms and simulation software that can exploit the possibilities of continuously evolving high-performance computing architectures and incorporate new insights from experimental and theoretical neuroscience is hence a fundamental activity in this field\cite{Morrison05a,Morrison08_267,Kunkel14_78,Jordan18_2,pronold2022,Pronold22_102952,Hines97,Stimberg19_47314,Yavuz16_18854}. 
It depends on large teams of software developers, who coordinate their work over years of development with many and fast update cycles.
Continuously monitoring the quality (correctness, usability, reproducibility) and the performance (speed, memory demands, energy costs) of the simulation code is mandatory\cite{Antz19,Saam2019}.
Without regularly testing the simulation code, the integration of new features or new optimizations may unknowingly lead to wrong results or performance breakdowns\cite{Albers22_837549}.
Detecting and understanding such unwanted behavior, comparing different hardware configurations or testing procedures, and sharing test results with other developers relies on an efficient tracking and organization of the corresponding benchmarking data and metadata.
\par
\begin{figure}[!htb]
  \centering
  \includegraphics[width=\textwidth]{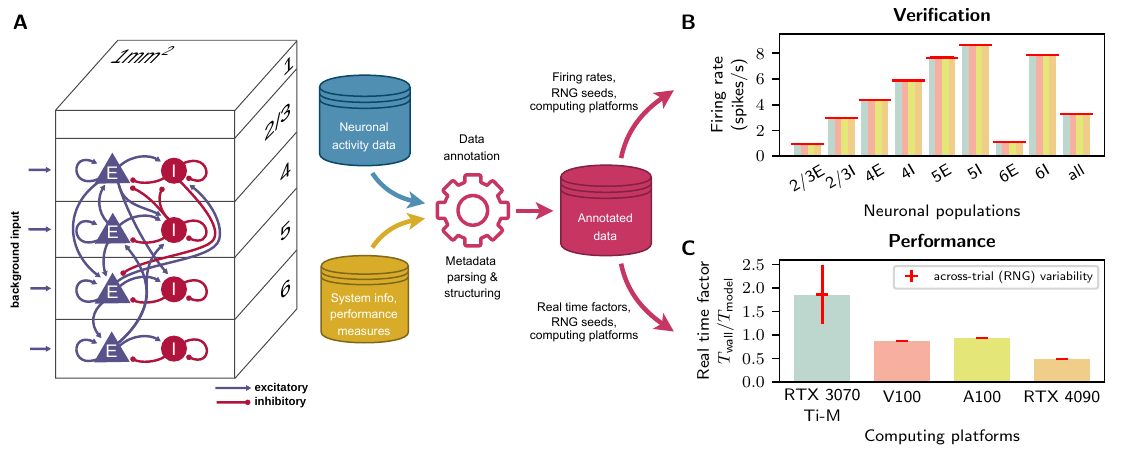}
  \caption{
    \textbf{Neuroscience use case.}
    \textbf{A)} 
    Left: Sketch of a model \cite{Potjans14_785} describing the activity dynamics generated by a local neuronal circuit of the mammalian neocortex (adapted from\,\cite{VanAlbada18_291}).
    The network model is composed of four excitatory (E; blue triangles) and four inhibitory neuronal populations (I; red circles); distributed across four cortical layers 2/3, 4, 5 and 6; and driven by background inputs.
    Neurons in the network are interconnected in a cell-type and layer specific manner (blue and red arrows).
    Right: The model generates neuronal activity data (``spikes'' and firing rates) as the primary neuroscientific data (blue cylinder).
    For each simulation instance, information about the model parameterization, random number generator (RNG) seeds, the hard- and software configuration, as well as the wall-clock times are stored in various files and formats as raw metadata (yellow cylinder).
    In a subsequent post-processing step (red gear), the metadata is parsed and structured by the \emph{Archivist}.
    The simulation data is annotated with this structured metadata and stored in a database for further usage (red cylinder).
    The database can flexibly be queried according to user interests (curved red arrows).         
    \textbf{B,\,C)}
    Verification (B) and performance benchmarking (C) as two exemplary types of data usage.
    B)
    Average activity level (firing rate) in each of the 8 neuronal populations 2/3E, \ldots, 6I depicted in panel A, obtained from simulations of the model on four different GPU platforms (see labels at horizontal axis in panel C).
    C)
    Real time factor (ratio between wall-clock time $T_\text{wall}$ and simulated biological time $T_\text{model}=10\,\text{s}$) for four different GPU computing platforms.
    Error bars (red) in B) and C) depict standard deviations across ten different model realizations (random-number generator [RNG] seeds) and simulation runs for each platform (error bars are partly too small to be visible).
  }
  \label{fig:usecase_neuro}
\end{figure}

The use case presented here demonstrates how the proposed metadata management practices and the \emph{Archivist} can help in fulfilling this task.
It illustrates a verification and performance benchmarking workflow for the neuronal-network simulation code  NEST GPU\cite{Golosio21_627620,Golosio23_9598} executed on a range of different hardware platforms (\cref{fig:usecase_neuro}).
It shows that the simulation code generates identical results on all tested platforms (\cref{fig:usecase_neuro}B), and highlights differences in the simulation speed (\cref{fig:usecase_neuro}C).
For illustration, we restrict this example to
\begin{itemize}
\item a specific test case: a model\cite{Potjans14_785} of a small piece of the mammalian cortex comprising about $80,000$ neurons and $300$ million synapses (\cref{fig:usecase_neuro}A),
\item a specific set of hardware platforms: four different GPU architectures (see axis labels in \cref{fig:usecase_neuro}B,C),
\item a specific verification metric: the average firing rates of neurons in different subpopulations of the network model (\cref{fig:usecase_neuro}B), and
\item a specific performance metric: the real time factor, i.e., the ratio between the wall-clock time and the simulated biological time \cref{fig:usecase_neuro}C).
\end{itemize}
Details on each of these aspects are given in \secref{sec:methods}. In a real performance-benchmarking setting, the same workflow would be (and has been \cite{Senk17_243,VanAlbada18_291,Jordan18_2,heittmann22_728460,Kurth_2022,Albers22_837549,Golosio23_9598,kauth23_1144143}) executed for a broader range of test cases, hardware configurations, verification metrics, and performance metrics.
\par
The entire pipeline underlying this use case is implemented using Snakemake\cite{Mlder2021} -- a lightweight yet powerful and flexible workflow manager.
Individual workflow components, such as the software deployment and compilation, the simulation execution, the storage of simulation data and raw metadata, the data and metadata post-processing, as well as the data exploration and presentation, correspond to specific Snakemake rules (for details, see \secref{sec:methods}).
During each execution of the simulation workflow, the (probabilistic) network model is simulated with $10$ different random realizations of the initial state and the network connectivity (different random-number-generator [RNG] seeds).
Each simulation instance generates single-neuron activity traces (spike trains) as well as population and time averaged firing rates as the primary neuroscientific data (blue cylinder in \cref{fig:usecase_neuro}A).
In addition, information about the model parameterization, RNG seeds, the hard- and software configuration, as well as the wall-clock times and real time factors are stored in various files and formats as raw metadata (yellow cylinder in \cref{fig:usecase_neuro}A).
After the simulation, the data and raw metadata are compressed, labeled with a unique identifier (uid), and uploaded to an instance of a Mongo database (MongoDB, \url{https://www.mongodb.com}, last access: 23 February 2024). 
During the metadata post-processing (red gear in  \cref{fig:usecase_neuro}A), the \emph{Archivist} uses specific parsers to extract the type of the computing platform, the RNG seed, and the real time factor from the raw metadata archives.
Subsequently, the extracted information is structured according to a user-specific template.
The link between the data and the metadata (data annotation) is established by attaching the data uid to the structured metadata.
At the end of the metadata post-processing, the structured metadata are uploaded to the database containing the data and raw metadata (red cylinder in \cref{fig:usecase_neuro}A).
The workflow is executed on four different GPU platforms, each sweep individually enriching the same database with its raw and structured metadata and annotated data.
The database containing the accumulated data and metadata from various instantiations of the simulation and metadata post-processing
workflow can be efficiently queried according to user interest (curved red arrows in \cref{fig:usecase_neuro}A).
For the verification and the performance assessment of the NEST GPU code, we extract the average firing rates and the real time factors, respectively, for each network realization (RNG seed) and computing platform, and plot the corresponding across-trial averages and standard deviations (\cref{fig:usecase_neuro}B,C).
\par
The source code underlying the workflow of this example can be found at \url{https://doi.org/10.5281/zenodo.13585723} (last access: 30 August 2024).

\phantomsection
\subsubsection*{Hydrology use case}
\setcurrentname{Hydrology use case}
\label{sec:hydro_examples}
The simulation of the hydrologic cycle is fundamental in environmental modeling. The movement and storage of water in the terrestrial system includes diverse processes. Key processes are the infiltration of precipitation into the subsurface, the storage of water in soils, and the removal of water from the soil via plant transpiration and subsurface runoff. A reliable estimation of water fluxes and storages is relevant for a wide range of sectors, like drinking water supply, agriculture, forestry, energy production, and transportation. Hydrologic models simulate hydrologic state variables and fluxes and their interaction. These models are created for different purposes according to the needs of specific sectors outlined above. These range from infrastructure planning for drinking water supply to drought quantification for agriculture and to climate projections for future water management. Hydrologic models represent fundamental processes like snow accumulation and melting, soil infiltration, evaporation and plant transpiration, surface and subsurface runoff, and river routing. Typical output variables are river discharge, evapotranspiration and soil moisture among others. Input variables are precipitation and air temperature. There are a wide range of computational workflows given the diverse purposes that hydrologic models are used for. Some of the most complex workflows are related to climate change projections where input variables are taken from different sources like climate model ensembles\cite{Samaniego2019}. Currently, metadata tracking in such workflows is often very limited and does not contain the comprehensive settings that are necessary to execute these workflows. 
\par
\begin{figure}[!htb]
  \centering
  \includegraphics[width=\textwidth]{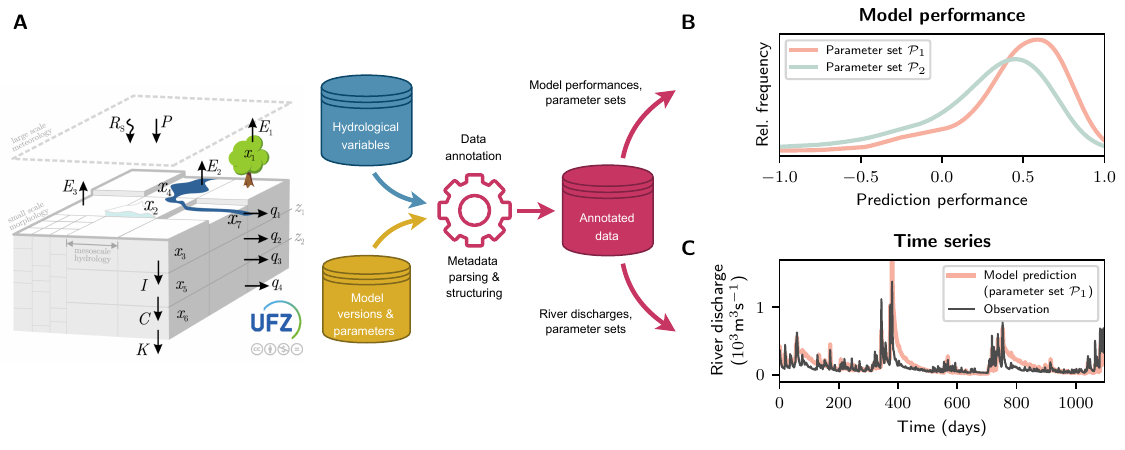}
  \caption{
    {\bf Hydrology use case.}
    \textbf{A)} A hydrologic model, here represented by the logo of the mesoscale Hydrologic Model (mHM \url{https://mhm-ufz.org}, last access: 25 June 2024), creates output (i.e., hydrological variables shown by the blue cylinder) and additional metadata information (yellow cylinder).
    A hydrologic model simulates the water cycle at the land surface.
    A typical output variable is river discharge (m$^3$s$^{-1}$).       
    Hydrological data is annotated by metadata post-processed by the \emph{Archivist}.
    \textbf{B)}
    Distribution of prediction performances of the hydrologic model across measurement stations for two parameter sets $\mathcal{P}_1$ and $\mathcal{P}_2$.
    Prediction performance is estimated by comparing simulated and observed river discharge for each measurement station.
    \textbf{C)} Time series of observed (black) and predicted river discharge (red; parameter set $\mathcal{P}_1$).
  }
  \label{fig:usecase_hydro}
\end{figure}

Here, we show how the proposed metadata management practices and the \emph{Archivist} can be applied to a workflow using a hydrologic model and which results can be derived from these (\cref{fig:usecase_hydro}).
A hydrologic model is visually represented by the logo of the mesoscale Hydrologic Model (mHM \cite{Samaniego2010, Kumar2013, Thober2019}) on the left in \cref{fig:usecase_hydro}A. Both output variables and metadata information, represented by the blue and yellow cylinder, respectively, are stored. In the hydrology use case, it is worth noting that the model output is orders of magnitude larger in terms of size than the associated metadata. The output variables are user-specified and contain a minimal set of metadata like unit and creation date.
The practices presented here  also allow us to create a set of additional metadata that provide comprehensive information about the workflow at hand (yellow cylinder in \cref{fig:usecase_hydro}).
For example, the entire configuration of the hydrologic model and its parameters can be recorded as well as the execution environment, version information, and all inputs and outputs. The \emph{Archivist} is then able to process the metadata information in a user-specific way. Two potential applications of the \emph{Archivist} are shown in \cref{fig:usecase_hydro}B and \cref{fig:usecase_hydro}C.
\par
In \cref{fig:usecase_hydro}B, a performance analysis of the hydrologic model is shown. A routine exercise in hydrologic modeling is parameter calibration.
This stems from the fact that hydrologic modeling follows the paradigm that model parameters govern model behavior. For example, water infiltration into the soil is controlled by a shape parameter that dictates how fast soil infiltration is decreasing when the soil is drying. This flexibility is fundamental to account for highly conductive soils (i.e., sandy soils) and low conductive soils (i.e., clay soil). Here, we are able to analyze how model performance depends on the particular choice of a parameter (\cref{fig:usecase_hydro}B). This can be done in principle over all the simulations that have ever been done using the hydrologic model, which is not possible without this comprehensive set of metadata. Performance is measured by comparing the model output, typically river discharge, with observations and calculating a performance measure from the simulated and observed time series. These performance measures are created in a way that the optimal value is at one and decreases to $-\infty$ for simulated and observed time series that are unrelated \cite{Gupta2009}.
Hydrologic models are run over a set of model parameters and given the performance of these, a cumulative distribution function can be calculated (\cref{fig:usecase_hydro}B)\cite{Thober2019, RakovecEtAlJ.Geophys.Res.Atmospheres2019}. \cref{fig:usecase_hydro}B shows that model configuration $\mathcal{P}_1$ is outperforming model configuration $\mathcal{P}_2$. Parameter set $\mathcal{P}_1$ leads to a larger frequency of performance around 0.5 and higher, while parameter set  $\mathcal{P}_2$ has larger relative frequency for lower performance values of zero and less.
Such an analysis can be done with all available parameter configurations, moreover subsets of parameters of interest can be created so that their performances are investigated. This allows users
to obtain a deeper understanding of how combinations of model parameters are affecting model performance at a level of detail that is currently not possible. 
\par
A critical task in the evaluation of hydrologic models is the visual inspection of the simulated and observed time series of a variable of interest.
\cref{fig:usecase_hydro}C depicts two time series of river discharge. The time series are based on simulations (red line in \cref{fig:usecase_hydro}C) and the black line is based on observations. The visual inspection then allows an understanding of the impact of the parameter calibration. More precisely, it allows an understanding of which part of the hydrograph is simulated well (e.g., high river discharge values, low river discharge values, or the transition phase between high and low values).
Having an annotated database at hand, as shown in \cref{fig:usecase_hydro}, allows users to freely select parameter configurations from already created simulations, not necessarily related to parameter optimization. Ongoing discussions on modern hydrologic sciences raise the need for reusability of hydrologic models results to better understand how process parametrizations are affecting model performance and behavior \cite{ClarkEtAlWaterResour.Res.2015}.
Such efforts largely benefit from enriching model results with metadata describing the simulation experiment. For example, discrepancies in model performance can be found to not only be due to differences in model parameters, but also model configuration.

\phantomsection
\section*{Discussion}
\setcurrentname{Discussion}
\label{sec:discussion}

Motivated by the expectation that rigorous metadata management is a prerequisite of reproducibility and interpretability of scientific results, this work proposes and applies practices for handling metadata in simulation research.
We derived a generic knowledge production workflow to illustrate that different types of metadata can be collected at different steps of a simulation, then be post-processed, and finally exploited alongside data.
To facilitate post-processing, we developed the \emph{Archivist} framework, a Python-based tool for parsing raw metadata files and combining the extracted information into a structured file.
The primary purpose of the tool is to attach it to existing workflows that require routine execution and do not incorporate other means of metadata tracking functionality yet. 
As example implementations, we presented a conceptual example and two use cases with varying degrees of complexity.
Our first use case was a hypothetical minimal example consisting of simple time-driven simulations where the collected metadata was used to track the performance of the simulation measured by the real time factor and the accuracy of the model obtained by analyzing the generated results.
Even if processing a single metadata entry is simple in nature, processing large volumes of metadata stemming from these simulations can prove challenging .
As the goal of the experiment was to find a compromise between performance and accuracy by exploring the parameter space, a consequently large amount of data and metadata was generated.
By automating metadata processing and structuring with the \emph{Archivist}, the experiment pipeline, from data generation to data usage, was streamlined.
Our second use case was a proof of concept of a benchmarking and verification workflow.
By running several simulations of the same model\cite{Potjans14_785} with NEST GPU\cite{Golosio21_627620}, the simulation performance was compared and simulation results were statistically verified across multiple hardware platforms.
Although the software setup was similar on all platforms, the format of the information retrieved directly from the hardware varied .
By leveraging the modularity of the \emph{Archivist}, parsing functions specific to each platform were implemented and could be exchanged without needing to modify the experiment workflow.
Because we uniformly structured the parsing output, the processed metadata describing different hardware platforms was homogenized and could be accessed the same way.
Our last use case was a routine parameter calibration procedure where multiple parameters for the mesoscale Hydrologic Model (mHM \cite{Samaniego2010, Kumar2013, Thober2019}) were evaluated across several simulation configurations.
Due to its routine nature, data and metadata for each procedure had accumulated over time. Furthermore, the collection of raw metadata is large even within a single calibration procedure, because information from each parameter set is collected.
The \emph{Archivist} helped combine and structure the metadata entries, allowing them to be stored in a database which could be queried for specific parameter sets to efficiently explore the results.

The current version of the \emph{Archivist} helps access and structure heterogeneous metadata collected from simulation workflows, yet it has limitations and room for improvement. 
With the \emph{Archivist's} modular design, described in \secref{sec:archivist}, users only need to define parsing functions to extract information from metadata entries, and use a data template to combine the results into desirable structures and formats.
Both parsing functions and data templates can be shared among groups to foster reusability of implemented workflows.
However, we note the need to properly document the defined functions and structures.
The same information can be interpreted differently by multiple users; one might mistakenly extract information with a different interpretation than expected.
A simple example would be reading a number with a large amount of decimal digits.
Functions can be implemented to use the exact number with all decimals, truncate after an arbitrary significance of decimal digits, or round to the nearest integer.
All of these implementations are valid interpretations and their use depends on how the information will be exploited later.
Appropriate documentation solves this problem by disambiguating implementations and increasing interpretability of the processed metadata.
Much like the source code and version of used models and simulators help describe a simulation experiment, the implementation and version of parsing functions describe the post-processing operations.
This is of particular importance considering that metadata processing for a given dataset is not a single final process, but can be refined for different needs.
By keeping a copy and sharing the raw metadata set, data can be subsequently annotated by the same or other users even if certain metadata entries were filtered out after an initial processing and annotation step.
Data is then further enriched by extending the available annotated information and can later be used for new studies.
Future contributions to the \emph{Archivist} could automatically interpret information in a data model to deduce the locations of files in a dataset and generate file target rules without the need for user definition
Given a sufficiently detailed description of dataset contents, parsing functions could also potentially be generated.

As mentioned in the Section titled \secref{sec:archivist}, there exist alternative tools for metadata post-processing, such as the AiiDA\cite{Huber20_4} parsers or DataLad\cite{Halchenko2021} metadata extractors.
AiiDA is a workflow manager that can use parsing plugins to extract information from files generated during the simulation workflow.
The parsed information can then be used as an output of a workflow step or attached as additional metadata for data provenance tracking.
DataLad is a distributed data management system where users can define extractors to parse files present in the database and extract metadata to annotate their respective datasets based on the data versioning system git-annex (\url{https://git-annex.branchable.com}, last access: 01 March 2024).
Conceptually, both these tools perform similar operations to the \emph{Archivist}: users define functions to extract information from files, these functions are used to parse the contents of the generated or stored data and extract metadata according to specific needs, and users can share these functions for further reusability.
The difference lies in how parsing is orchestrated and how the extracted metadata is handled.
In both AiiDA and DataLad, parsing is triggered on a file-by-file basis.
Although the metadata can be integrated into a knowledge base (i.e., the provenance graph for AiiDA or as annotated information for DataLad), each piece of parsed information is treated independently and further processing is required to combine it into a comprehensive output.
In contrast, the parsers employed by the \emph{Archivist} operate on an arbitrary number of files automatically determined by user-defined rules.
Furthermore, the \emph{Archivist} empowers users with the possibility to define a custom data template based on JSON Schema to structure and combine the parsed metadata.
The ability to use a custom data template helps ensure that previous and future post-processing results are consistently structured. 
Because our practices promote storing unprocessed workflow results as well as processed results, previous results can be reprocessed with a newer data template if a different structure is needed for a new purpose, thereby fostering reusability.
Additionally, users can base their template on vocabularies or schemas such as RO-Crate\cite{SoilandReyes2022}, CodeMeta, or Bioschemas\cite{bioschemas} for further interoperability of their post-processing results. 
We note that our intention is not to create a replacement for features in AiiDA or DataLad, but to implement a standalone tool that can transparently suit specific user needs.
Dedicated workflow manager tools such as Snakemake\cite{Mlder2021} or AiiDA are particularly useful for implementing a completely new workflow - in addition to increased reusability and interoperability of workflows, these tools offer software environment handling, HPC cluster job handling, and data provenance tracking among other features\cite{Diercks23}.
Data provenance in particular is useful for describing the contents of data generated during simulations.
Rather than being incompatible with our framework, these alternative tools could actually be combined with it by using the \emph{Archivist} as a parsing and structuring backend.
This combined implementation would feature automated experiment description, flexible data characterization, and efficient data annotation and exploration. 
To this end, future work could be done on a combined implementation of the \emph{Archivist} with existing tools for automated provenance tracking and enhanced storage platforms.

We note that, although our abstract simulation workflow \cref{fig:workflow} (blue) can represent a wide range of real-world  workflows, it is certainly not an absolute representation.
This implies that our metadata management practices defined through this abstract workflow are not fully applicable to all workflows.
Very intricate workflows, for example, may deal with challenges that do not allow for a one-off intervention of incorporating the \emph{Archivist}, but call for further customized solutions instead.
Particular examples are highly evolving workflows containing steps that are frequently swapped, added, or removed requires keeping track of which metadata to collect and updating the corresponding post-processing methods.
Even updating only the simulation workflow is time-consuming, so updating the metadata post-processing workflow too will require even more effort.
Furthermore, simulation workflows that require complex dynamical inputs prepared in a separate preceding workflow, or that incorporate separate subsequent post-processing workflows -- if metadata on these inputs or subsequent workflows is not collected during execution, they would be irretrievably lost.
A final example where our practices may not be fully applicable are simulations where the size of produced data and/or metadata is too large to store both in ``raw'' and processed format.
These workflows require that data and metadata are processed, possibly filtered, and annotated during runtime before being able to be stored.
Decoupling the post-processing of metadata describing data generated during the simulation experiment is not beneficial in these cases and would most likely lead to inconsistency issues.
Nonetheless, even if not fully applicable, our practices can serve as a stepping stone for these complex workflows as metadata describing scientific experiments remain essential for long-term storage and sharing results.

Our proposed practices for handling metadata in simulation workflows are applicable to a wide range of scientific disciplines with domain-specific workflows and metadata conventions.
New workflows can benefit from these practices to ensure that metadata is handled accordingly.
The practices can also be implemented in existing workflows without major restructuring of established code.
In this case, our proof of concept tool, the \emph{Archivist}, provides a flexible solution for parsing and structuring heterogeneous metadata files.
Through the use cases presented in this work we illustrate how our practices and tool support sustainable numerical workflows, fostering reproducibility and data reuse in simulation-based research.

\phantomsection
\section*{Methods}
\setcurrentname{Methods}
\label{sec:methods}

\phantomsection
\subsection*{Details on the neuroscience use case}
\setcurrentname{Details on the neuroscience use case}
\label{sec:neuro_use_case_details}
The \secref{sec:neuroscience_example} focuses on a workflow for benchmarking and verification of a spiking neural network simulator across multiple hardware platforms.
The workflow implementation can be found at \url{https://doi.org/10.5281/zenodo.13585723} (last access: 30 August 2024).

\paragraph{NEST GPU.}
The simulator in question is NEST GPU\cite{Golosio21_627620,Golosio23_9598}, an open library for simulation of large-scale networks of spiking neurons, written in the C++ and CUDA-C++ programming languages (source code: \url{https://github.com/nest/nest-gpu}, last access: 23 February 2024; documentation: \url{https://nest-gpu.readthedocs.io}, last access: 23 February 2024).

\paragraph{Hardware platforms.}
The goal of this experiment is to determine whether the simulator would produce the same results across different hardware platforms.
As the models simulated rely on floating point calculations, artifacts during numerical computations may arise during simulation and could potentially accumulate to observable differences in produced results.
In particular, GPU architectures can vary highly between generations, which increases the chances of such divergence.
For this reason we test the simulator with different GPUs models, both consumer and data center level, with varying architectures: one laptop with the consumer GPU RTX 3070 Ti Mobile with CUDA version 12.2; two compute clusters, JUSUF \cite{VonStVieth2021} and JURECA-DC \cite{Thoernig2021}, both using CUDA version 11.3 and equipped with the data center GPUs V100 and A100, respectively; and a workstation with the consumer GPU RTX 4090 with CUDA version 11.4.

\begin{table}[!htb]
\caption{Hardware configuration of the different platforms used.. Cluster information is given on a per node basis.\label{tab:hardware_conf}}
\newcolumntype{C}{>{\centering\arraybackslash}X}
\begin{tabularx}{\textwidth}{CCC}
\toprule
\textbf{System}	& \textbf{CPU}	& \textbf{GPU}\\
\midrule
Laptop	& Intel Core  i7-12800H vPro, 14 cores (6 P-cores + 8 E-cores), P-core 4.8GHz / E-core 3.7GHz	& NVIDIA RTX 3070 Ti Mobile\textsuperscript{2}, 1410 MHz, 8 GB GDDR6, 5888 CUDA cores  \\
\midrule
JUSUF cluster	& 2× AMD EPYC 7742, 2× 64 cores, 2.25 GHz	& NVIDIA V100\textsuperscript{1}, 1530 MHz, 16 GB HBM2e, 5120 CUDA cores \\
\midrule
JURECA-DC cluster	& 2× AMD EPYC 7742, 2× 64 cores, 2.25 GHz	& NVIDIA A100\textsuperscript{2}, 1410 MHz, 40 GB HBM2e, 6912 CUDA cores \\
\midrule
Workstation	& Intel Core  i9-10940X, 14 cores, 3.30 GHz	& NVIDIA RTX 4090\textsuperscript{3}, 2520 MHz, 24 GB GDDR6X, 16384 CUDA cores  \\
\bottomrule
\end{tabularx}
\noindent{\footnotesize{\textsuperscript{1} Volta architecture: \url{https://developer.nvidia.com/blog/inside-volta}, last access: 25 June 2024}}\\
\noindent{\footnotesize{\textsuperscript{2} Ampere architecture: \url{https://developer.nvidia.com/blog/nvidia-ampere-architecture-in-depth}}, last access: 25 June 2024}\\
\noindent{\footnotesize{\textsuperscript{3} Ada Lovelace architecture: \url{https://www.nvidia.com/en-us/technologies/ada-architecture/}, last access: 25 June 2024}}
\end{table}

\paragraph{Network model.}
The model used to evaluate the simulator is the cortical microcircuit of Potjans and Diesmann\cite{Potjans14_785} which represents a 1 mm$^2$ patch of early sensory cortex at the biological plausible density of neurons and synapses (depicted in \cref{fig:usecase_neuro}A).
The model was simulated previously on different simulation platforms\cite{VanAlbada18_291,Golosio21_627620,Knight21_15}, and in a recent study dynamics and performance were evaluated using NEST GPU\cite{Golosio23_9598}.
The availability of both dynamics and performance data make this model ideal for verification tests. Furthermore, the inherent complexity and scale of the network are good candidates to confirm whether numerical artifacts would be generated even on data center level GPUs.

\paragraph{Verification metrics.}
To verify the simulated dynamics, we collect spiking activity of all neuron populations of the model, and compute their average firing rate.
Simulations are performed using a time step of $0.1$\,ms and $500$\,ms of network dynamics are simulated before recording spiking activity to avoid transients.
Then, we record spiking activity of the subsequent $10$\,s of network dynamics.
As shown in previous studies\cite{Dasbach21_90}, the average firing rate of a population is computed as the number of recorded spikes emitted by all neurons in the population, averaged by the number of neurons in the population, and normalized by the duration of the recording.

\paragraph{Performance metrics.}
To measure the performance of the simulation in each hardware platform we use the \textit{real time factor} as defined in \secref{sec:minimal_example}.
Here we use internal timers included in the model definition to compare the time needed for state propagation of model dynamics and the simulated biological time.

\paragraph{Data generation.}
As a proof of concept, we devised a minimal workflow where we follow our previously defined practices on metadata collection, post-processing, and annotation to populate a database.
Here, we describe the steps for software preparation, simulation execution, and metadata collection (see \cref{fig:workflow}).
For simplicity, we assume that all software dependencies for the simulator and for the data processing pipeline are already present in the target platform.
We also assume that no job schedulers are needed for execution.
The workflow, implemented using Snakemake,  consists of nine rules underlying the production and storage of raw simulation data and metadata, two rules for metadata post-processing, and one rule for data exploration and usage.
This implementation was designed so each rule can be dynamically configured through a dedicated file to increase flexibility and reusability for different simulation scenarios (such as different simulators and models).
The workflow starts by cloning two separate repositories, one for the simulator and another for the model.
Following this, the simulator is compiled using its CMake (\url{https://cmake.org}, last access: 23 February 2024) installation infrastructure.
Then the model is consecutively run in a sequence of independent simulations each with different random number generation seed.
The spiking activity predicted by the model is recorded and constitutes the primary data output of each simulation.
Special care is also taken to monitor the simulated biological time and the wall-clock time for performance data.
Information on the system environment before and after job execution is recorded through a dedicated metadata collection script.
This allows us to compare the state of the system before and after setting up the simulation environment.
Additional metadata produced before execution, such as configuration files, execution scripts, compilation output, shell output logs, and the internal metadata tracked by Snakemake itself, are stored as raw metadata.
After each simulation run and metadata collection, a post-processing step computes the population-averaged firing rates from the recorded spike data (\cref{fig:usecase_neuro} B), as well as the  real time factor (\cref{fig:usecase_neuro} C).
Finally, the produced data and metadata are compressed into archives.
As storage platform, we use an instance of a MongoDB.
By employing its file storage system GridFS (\url{https://www.mongodb.com/docs/manual/core/gridfs}, last access 23 February 2024), we can upload the compressed archives and get a unique identifier for each.

\paragraph{Vocabulary usage example.}
In Listing~\ref{lst:vocab_example} we show a partial example output of a metadata post-processing pipeline.
In this example, input and output file names were extracted from the workflow configuration file and formatted according to the Bioschemas vocabulary (\url{https://bioschemas.org/types/ComputationalWorkflow/1.0-RELEASE}, last visited: 23 January 2025).
If specific file names are changed in the configuration this update would be reflected in the output of a subsequent metadata post-processing pipeline.
These were then used to complete a simple Bioschemas document \texttt{bioschemas.jsonld} included with the workflow source code.
Although valid, the document content is minimal and only shown as a proof of feasibility.
Users would naturally employ the full capabilities of Bioschemas to describe their own workflows. 

\begin{lstlisting}[caption={Partial output of an \emph{Archivist} post-processing pipeline. Extracted information from workflow configuration file was formatted according to the Bioschemas vocabulary.},label=lst:vocab_example]
{
    "inputs": [
        {
            "@type": "FormalParameter",
            "dct:conformsTo": "https://bioschemas.org/profiles/FormalParameter/1.0-RELEASE",
            "name": "ngpu_config",
            "encodingFormat": "text/json",
            "valueRequired": true,
            "description": "Configuration file of workflow."
        },
        ...
    ],
    "outputs": [
        ...
        {
            "@type": "FormalParameter",
            "dct:conformsTo": "https://bioschemas.org/profiles/FormalParameter/1.0-RELEASE",
            "name": "structured_metadata",
            "encodingFormat": "text/json",
            "description": "Output file of metadata post-processing."
        },
        ...
    ]
}
\end{lstlisting}

\phantomsection
\subsection*{Details on the hydrology use case}
\setcurrentname{Details on the hydrology use case}
\label{sec:hydro_use_case_details}

The \secref{sec:hydro_examples} presents how the metadata \emph{Archivist} can be applied in the hydrological modeling sciences.

\paragraph{mHM.}
 An essential tool of this use case is the employed mesoscale hydrologic model mHM \cite{Samaniego2010, Kumar2013, Thober2019} (\url{https://mhm-ufz.org}, last access: 21 March 2025). mHM is a grid-based, spatially distributed hydrological model driven by daily precipitation, temperature, and potential evapotranspiration. It accounts for major hydrological processes such as snow accumulation and melt, canopy interception, soil infiltration, evapotranspiration, deep percolation, baseflow generation, and river routing. The open-source model code repository is available and is under active development and maintenance (\url{https://git.ufz.de/mhm/mhm}, last access: 21 March 2025).
A general overview on the model processes and parameterization can be obtained from \cite{Samaniego2010} and \cite{Kumar2013}. The model is an integral part of the German Drought Monitor (\url{https://www.ufz.de/duerremonitor}, last access: 21 March 2025). mHM was also applied and evaluated in multiple climatological regions, including Europe \cite{Thober2015, Rakovec2016}, West Africa\cite{Dembl2020}, India\cite{Saha2021}, and the conterminous United States\cite{Livneh2015,RakovecEtAlJ.Geophys.Res.Atmospheres2019}.

\paragraph{Hardware platforms.}
mHM simulations were carried out on the Computing Cluster EVE, a joint effort of both the Helmholtz Centre for Environmental Research - UFZ (\url{http://www.ufz.de/}, last access: 21 March 2025) and the German Centre for Integrative Biodiversity Research (iDiv) Halle-Jena-Leipzig (\url{https://www.idiv.de/}, last access: 25 June 2024). The main compute hardware of EVE comprises a) 42 compute nodes with dual socket Intel Xeon 6348 CPUs with 512 Gigabytes of DDR4 main memory, two of these also include NVIDIA Tesla A100 GPGPUs, and b) 27 compute nodes with dual socket Intel Xeon Gold 6148 CPUs with up to 1536 Gigabytes of DDR4 main memory, two of which include NVIDIA Tesla V100 GPGPUs. The central network component of the cluster is an Intel Omni-Path 100 Series high performance interconnect, providing all compute nodes with non-blocking EDR bandwidth (100 Gigabit per second). All compute nodes share a 4.5 Petabyte IBM Spectrum Scale file system. The system performed with over 164 teraFLOPS under a High-Performance LINPACK (HPL) benchmark. For the simulations of this project, we have used CPU cores exclusively.

\paragraph{Model setup.}
The mHM model is executed within the test basin provided along with the model source code, the so called ``test basin". The test basin coming with the mHM source code is for the Moselle River basin upstream of Perl, a place, where the Moselle River leaves France and enters Luxembourg and Germany (Moselle Basin). The catchment area is approximately $11,500\text{ km}^2$, altitude ranges between 150 and 1300 m. a.m.s.l. The Moselle River originates from the Vosges Mountains and is a tributary of the Rhine River. The origin of data used in the test example is provided on \url{https://mhm-ufz.org/docs/} (last access: July 18th, 2024).

\paragraph{Performance metrics.}
The hydrology use cases makes use of performance metrics that are commonly used in hydrologic modeling. Explicitly, it is the Kling-Gupta efficiency (KGE \cite{Gupta2009}), that can be used to compare two time series. The metric combines the long-term mean, long-term variance, and Pearson correlation coefficient.

\paragraph{Data generation.}
The mHM simulations have been facilitated using the ecFlow workflow manager \cite{Bahra2011} developed at the European Centre for Medium-Range Weather Forecasts (ECMWF). ecFlow is a client/server workflow package that allows users to execute any number of simulations. It is tailored to work on HPCs and allows to easily restart workflows if hardware and software failures occur. We have created a simple suite with three tasks. The first task is the compilation of mHM using CMake (\url{https://cmake.org}, last access: 23 February 2024). The second tasks executes the hydrologic model. The third task is a post-processing step that creates plots like the ones shown in \cref{fig:usecase_hydro}B and~C. We have now also added the \emph{Archivist} to these workflows to manage the metadata.

\phantomsection
\section*{Data Availability}
\setcurrentname{Data Availability}
\label{sec:data_availability}
No external dataset or input data were used for this study.

\phantomsection
\section*{Code Availability}
\setcurrentname{Code Availability}
\label{sec:code_availability}
Source code, example implementations, and additional details of the \emph{Archivist} framework can be found at \url{https://doi.org/10.5281/zenodo.13442425} (last access: 30 August 2024).
An implementation of the metadata post-processing performed for the \secref{sec:minimal_example} can be found at \url{https://doi.org/10.5281/zenodo.13442425} (last access: 30 August 2024).
The workflow implementation for the \secref{sec:neuroscience_example} as well as the data and script required to create \cref{fig:usecase_neuro} can be found at \url{https://doi.org/10.5281/zenodo.13585723} (last access: 30 August 2024).
The model implementation of the \secref{sec:hydro_examples} used to generate the data required to create \cref{fig:usecase_hydro} is available at \url{https://doi.org/10.5281/zenodo.1069202} (last access: 26 January 2025).

\phantomsection
\section*{Author Contributions}
\setcurrentname{Author Contributions}
\label{sec:contributions}
All authors contributed to the conceptual metadata management framework, and the design of the example use cases.
J.V.~and M.K.~implemented the metadata \emph{Archivist}.
J.V.~implemented the neuroscience use case.
M.K.~implemented the hydrology use case.
All authors wrote and reviewed the manuscript.

\phantomsection
\section*{Competing Interests}
\setcurrentname{Competing Interests}
\label{sec:interests}

The authors declare no competing interests.

\phantomsection
\section*{Acknowledgments}
\setcurrentname{Acknowledgments}
\label{sec:ack}
The authors thank
Dennis Terhorst, and
Guido Trensch
for constructive discussions.
This project has received funding
from the Initiative and Networking Fund of the Helmholtz Association in the framework of the Helmholtz Metadata Collaboration (HMC) project call (ZT-I-PF-3-026)
and under project number SO-092 (Advanced Computing Architectures, ACA),
the Helmholtz Joint Lab ``Supercomputing and Modeling for the Human Brain'',
the European Union’s Horizon Europe Programme under the Specific Grant Agreement No. 101147319 (EBRAINS 2.0 Project),
and HiRSE\_PS, the Helmholtz Platform for Research Software Engineering - Preparatory Study, an innovation pool project of the Helmholtz Association.
The authors gratefully acknowledge the computing time granted by the JARA Vergabegremium and provided on the JARA Partition part of the supercomputer JURECA at Forschungszentrum J\"ulich (computation grant JINB33).
They further acknowledge the use of Fenix Infrastructure resources, which are partially funded from the European Union's Horizon 2020 research and innovation programme through the ICEI project under the grant agreement No.\,800858.
The work was carried out in part within the HMC Hub Information at the Forschungszentrum J\"ulich.

\end{document}